# Primer on the Gene Ontology

Pascale Gaudet[1,*], Nives Škunca[2,3,4], James C. Hu[5], and Christophe Dessimoz[4,6]

[1]SIB Swiss Institute of Bioinformatics, 1 Michel-Servet, 1211 Geneva, Switzerland

[2]ETH Zurich, Computer Science, Universitätstr. 19, 8092 Zurich, Switzerland

[3]SIB Swiss Institute of Bioinformatics, Universitätstr. 19, 8092 Zurich, Switzerland

[4]University College London, Gower St, London WC1E 6BT, UK

[5]Department of Biochemistry and Biophysics, Texas A&M University and Texas AgriLife Research, College Station, TX USA

[6]Department of Ecology and Evolution and Center of Integrative Genomics, University of Lausanne, Biophore, 1015 Lausanne, Switzerland

*Corresponding author: pascale.gaudet@isb-sib.ch

**Abstract**

The Gene Ontology (GO) project is the largest resource for cataloguing gene function. The combination of solid conceptual underpinnings and a practical set of features have made the GO a widely adopted resource in the research community and an essential resource for data analysis. In this chapter, we provide a concise primer for all users of the GO. We briefly introduce the structure of the ontology and explain how to interpret annotations associated with the GO.

Keywords. Gene Ontology structure, evidence codes, annotations, gene association file (GAF), GO files, function, vocabulary, annotation evidence

## 1. Introduction

The key motivation behind the Gene Ontology (GO) was the observation that similar genes often have conserved functions in different organisms ***(1)***. Clearly, a common vocabulary was needed to be able to compare the roles of orthologous genes (and their products) across different species. The value of comparative studies of biological function across systems predates Jacques Monod's statement that "anything found to be true of *E. coli* must also be true of elephants" ***(2)***. The Gene Ontology aims to produce a rigorous shared vocabulary to describe the roles of genes across different organisms ***(1)***. The GO project consists of the *Gene Ontology* itself, which models biological aspects in a structured way, and *annotations*, which associate genes or gene products with terms from the Gene Ontology. Combining information from all organisms in one central repository makes it possible to integrate knowledge from different databases, to infer the functionality of newly discovered genes, and to gain insight into the conservation and divergence of biological subsystems.

In this primer, we review the fundamentals of the GO project. The chapter is organised as answers to five essential questions: What is the GO? Why use it? Who develops it and provides annotations? What are the elements of a GO annotation? And finally, how can the reader learn more about GO resources?

## 2. What is the Gene Ontology?

The Gene Ontology is a controlled vocabulary of terms to represent biology in a structured way. The terms are subdivided in three distinct ontologies that represent different biological aspects: Molecular Function (MF), Biological Process (BP), and Cellular Component (CC) ***(1)***. These ontologies are non-redundant and share a common space of identifiers and a well-specified syntax.

Terms are linked to each other by relations to form a hierarchical vocabulary (add ref to Janna Hasting's chapter). This is often modelled as a graph in which the relationships form the directed edges, and the terms are the nodes (Figure 1). Since each term can have multiple relationships to broader parent terms and to more specific child terms, the structure allows for more expressivity than a simple hierarchy.

The full GO is large: in October 2015, the full ontology specification had 43835 terms, 73776 explicitly encoded `is_a` relationships, 7436 explicitly encoded `part_of` relationships, and 8263 explicitly encoded `regulates`, `negatively_regulates` or `positively_regulates` relationships. This level of detail is not necessary for all applications. Many research groups who do GO annotations for specific projects use the generic GO-slim file, which is a manually curated subset of the Gene Ontology containing general, high-level terms across all biological aspects. There are several GO slims[1], ranging from the general Generic GO slim developed by the GO Consortium to more specific ones, such as the Chembl Drug Target slim (http://wwwdev.ebi.ac.uk/chembl/target/browser) .

---

[1] http://geneontology.org/page/go-slim-and-subset-guide

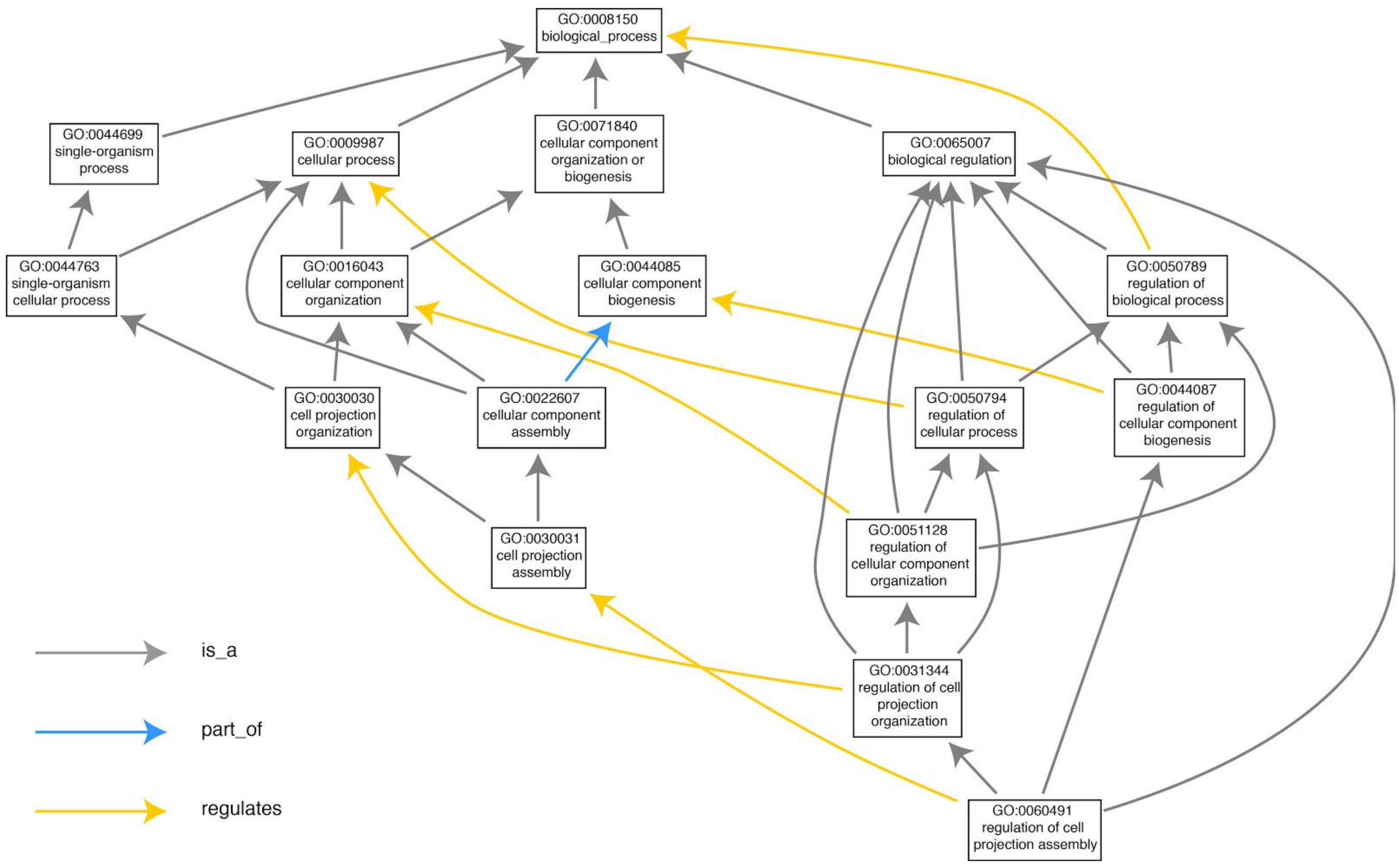

**Figure 1:** The structure of the Gene Ontology (GO) is illustrated on a subset of the paths of the term "regulation of cell projection assembly," GO:0060491, to its root term. The GO is a directed graph with terms as nodes and relationships as edges; these relationships are either `is_a`, `part_of`, `has_part`, or `regulates`. In its basic representation, there should be no cycles in this graph, and we can therefore establish parent (more general) and child (more specific) terms (see Chap. XX for more details on the different representations; cross-reference to Moni's chapter). Note that it is possible for a term to have multiple parents. This figure is based on the visualization available from the AmiGO browser, generated on November 6, 2015. ***(3)***.

To keep up with the current state of knowledge, as well as to correct inaccuracies, the GO undergoes frequent revisions: changes of relationships between terms, addition of new terms, or term removal (obsoletion). Terms are never deleted from the ontology, but their status changes to obsolete and all relationships to the term are removed ***(4)***. Furthermore, the name itself is preceded by the word "obsolete" and the rationale for the obsoletion is typically found in the Comment field of the term. An example of an obsolete term is GO:0000005, "obsolete ribosomal chaperone activity." This MF GO term was made obsolete "because it refers to a class of gene products and a biological process rather than a molecular function"[2]. Changes to the *relationships* do not impact annotations, because annotations are associated with a given GO term regardless of its relationships to other terms within the GO. Obsoletion of terms however have an impact on *annotations* associated with them: in some cases, the old term can be automatically replaced by a new or a parent one; in others, the change is so important that the annotations must be manually reviewed.

However, these changes can affect the analyses done using the ontology. In articles or reports, it is good practice to provide the version of the file used for a particular analysis. In GO, the version number is the date the file was obtained from the GO site (GO files are updated daily).

**3. Why use the Gene Ontology?**

Because it provides a standardised vocabulary for describing gene and gene product functions and locations, the GO can be used to query a database in search of genes' function or location within the cell or to search for genes that share characteristics ***(5)***. The hierarchical structure of the GO allows to compare proteins annotated to different terms in the ontology, as long as the terms have relationships to each other. Terms located close together in the ontology graph (i.e., with a few intermediate terms between them) tend to be semantically more similar than those further apart (see Chap. of Catia Pesquita on comparing terms).

The GO is frequently used to analyse the results of high-throughput experiments. One common use is to infer commonalities in the location or function of genes that are over- or under-expressed ***(4, 6)*** [+cross-reference to Sebastian Bauer's chapter]. In functional profiling, the GO is used to determine which processes are different between sets of genes. This is done by using a likelihood-ratio test to determine if GO terms are represented differently between the two gene sets ***(4)***.

[2] https://www.ebi.ac.uk/QuickGO/GTerm?id=GO:0000005

Additionally, the GO can be used to infer the function of unannotated genes. Gene predictions with significant similarity to annotated genes can be assigned one or several of the functions of the characterized genes. Other methods such as the presence of specific protein domains can also be used to assign GO terms ***(7, 8)***. This is discussed in Chap. XX (x-ref to Cozzetto and Jones).

A wealth of tools—web-based services, standalone software, and programing interfaces—has been developed for applying the GO to various tasks. Some of these are presented in Chap. XXX (x-ref to Moni Munoz-Torres's chapter).

While Gene Ontology resources facilitate powerful inferences and analyses, researchers using the GO should familiarise themselves with the structure of the ontology and also with the methods and assumptions behind the tools they use to ensure that their results are valid. Common pitfalls and remedies are detailed in Chap. XX (x-ref to Gaudet and Dessimoz chapter).

## 4. Who develops the GO and produces annotations?

The GO Consortium consists of a number of large databases working together to define standardised ontologies and provide annotations to the GO ***(9)***. The groups that constitute the GO consortium include UniProt ***(10)***, Mouse Genome Informatics ***(11)***, *Saccharomyces* Genome Database ***(12)***, Wormbase ***(13)***, Flybase ***(14)***, dictyBase ***(15)***, and TAIR ***(16)***. In addition, several other groups contribute annotations, such as EcoCyc ***(17)*** and the Functional Gene Annotation group at University College London ***(18)***[3]. Within each group, biocurators assign annotations according to their expertise ***(19)***. Further, the GO Consortium has mechanisms by which members of the broader community (see chapter on Community Annotations) can suggest improvements to the ontology and annotations.

[3] Full list at http://geneontology.org/page/go-consortium-contributors-list

**5. What are the elements of a GO annotation?**

This section describes the different elements composing an annotation and some important considerations about each of them. The annotation process from a curator standpoint is discussed in detail in the chapter by Gaudet and Poux (cross-reference).

Fundamentally, a GO annotation is the association of a gene product with a GO term. From its inception, the GO Consortium has recognized the importance of providing supporting information alongside this association. For instance, annotations always include information about the evidence supporting the annotation.

Over time, the GO Consortium standards for storing annotations have evolved to improve this representation. Annotations are now stored in one of two formats: GAF (Gene Association File), and the more recent GPAD (Gene Product Association Data). The two formats contain the same information but there are differences in how the data is normalised and represented (discussed in more details in Chap. XXX, x-ref to Monica Munoz-Torres's chapter). In this primer, we focus on the former. The representation of an annotation in the GAF file format 2.1 is shown in Figure 2. It contains 17 fields (also sometimes referred to as "columns"). We describe them in this section.

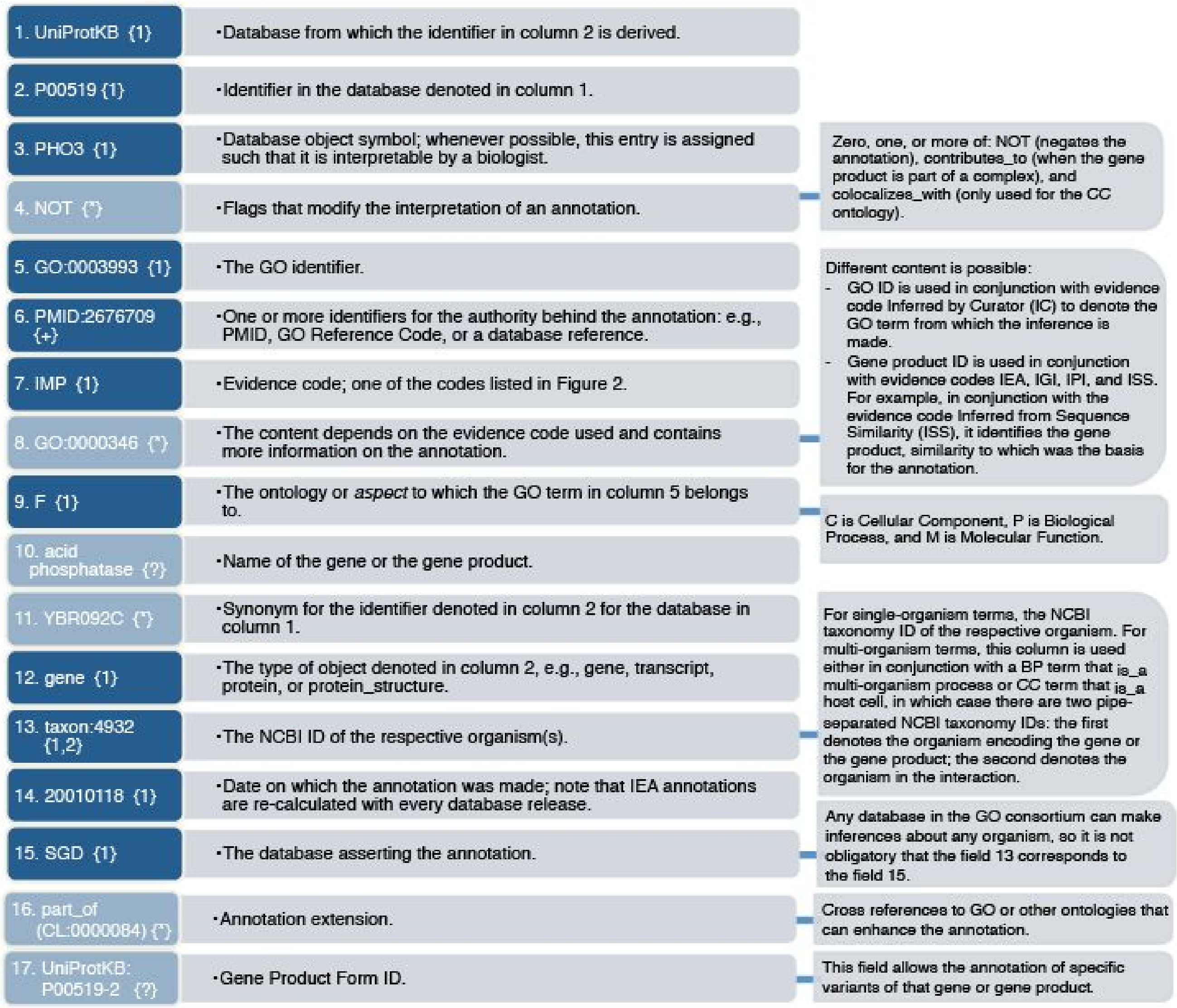

**Figure 2. Gene Association File (GAF) 2.1 file format described with example elements.** In the GAF file, each row represents an annotation, consisting of up to 17 tab-delimited fields (or columns). This figure describes these fields in the order in which they are found in the GAF file. Light blue colour denotes non-mandatory fields, and these are allowed to be empty in the GAF file. The cardinality—the number of elements in the field—is denoted with the symbol(s) in curly brackets: {?} indicates cardinality of zero or one; {*} indicates that any cardinality is allowed; {+} indicates cardinality of one or more; {1} indicates that cardinality is exactly one; {1,2} indicates that cardinality is either one or two. When cardinality is greater than 1, elements in the field are separated with a pipe character or with a comma; the former indicates 'OR' and the latter indicates 'AND'. The GO term assigned in column 5 is always the most specific GO term possible.

### 5.1 Annotation object

The annotation object is the entity associated with a GO term— a gene, a protein, a non-protein-coding RNA, a macromolecular complex, or another gene product. Seven fields of the GAF file specify the annotation object. Each annotation in the GO is associated with a database (field 1) and a database accession number (field 2) that together provide a unique identifier for the gene, the gene product, or the complex. For example, the protein record P00519 is a database object in the UniProtKB database (Figure 2). The database object symbol (field 3), the database object name (field 10), and the database object synonyms (field 11) provide additional information about the annotation object. The database object type specifies whether the object being annotated is a gene, or a gene product (e.g., protein or RNA; field 12). The organism from which the annotation object is derived is captured as the NCBI taxon ID (taxon; field 13); the corresponding species name can be found at the NCBI taxonomy website[4].

GO allows capturing isoform-specific data when appropriate, for example UniProtKB accession numbers P00519-1 and P00519-2 are the isoform identifiers for isoform 1 and 2 of P00519. In this case, the database ID still refers to the main isoform, and an isoform accession is included in the GAF file as "Gene Product Form ID" (field 17).

### 5.2 GO term, annotation extension, and qualifier

Three fields are used to specify the function of the annotation object. Field 5 specifies the GO term, while field 9 denotes the sub-ontology of GO, either Molecular Function, Biological Process, or Cellular Component. While this information is also encoded in the GO hierarchy, explicitly denoting the sub-ontology allows simplifies parsing of the annotations according to the GO aspect. Field 4 denotes the qualifier. One of three qualifiers can modify the interpretation of an annotation: "`contributes_to`", "`colocalizes_with`" and "`NOT`." This field is not mandatory, but if present it can profoundly change the meaning of an annotation ***(4)***. Thus, while the *producers* of annotations may omit qualifiers, applications that *consume* GO annotations must take them into account. The importance of qualifiers is discussed in more detail in Chapter XX (Gaudet and Dessimoz).

---

[4] http://www.ncbi.nlm.nih.gov/taxonomy

An additional field, field 16, is a recent addition to combine more than one term or concept (protein, cell type, etc.) in the same annotation. For example[5], if a gene product Slp1 is localized to the plasma membrane of T-cells, the GAF file field 16 would contain the information "part_of(CL:0000084 T cell)." Here, CL:0000084 is the identifier for T-cell in the OBO Cell Type (CL) Ontology. This is covered in details in Chap. XX (x-ref to Lovering and Huntley's chapter on annotation extensions).

### 5.3 Evidence code and reference field

Three fields in the GAF file describe the evidence used to assert the annotation: the Reference (field 6), the Evidence Code (field 7), and the With/From (field 8). The Evidence Code informs the type of experiment or analysis that supports the annotation. There are 21 evidence codes, which can be grouped in three broad categories: experimental annotations, curated non-experimental annotations, and automatically assigned (also known as electronic) annotations (Figure 3). The Reference field specifies more details on the source of the annotation. For example, when the evidence code denotes an experimentally supported annotation, the Reference will contain the PubMed accession ID (or a DOI if no PubMed ID is available) of the journal article which underpins the annotation, or a GO_REF identifier that refers to a short description of the assignment method, accessible on the GO website[6]. When the evidence code denotes an automatically assigned annotation, i.e. IEA, the reference will contain a GO_REF identifiers that specify more details on the automatic assignment, e.g., annotation via the InterPro resource ***(20)***.

---

[5] http://wiki.geneontology.org/index.php/Annotation_Extension#The_basic_format
[6] http://www.geneontology.org/cgi-bin/references.cgi

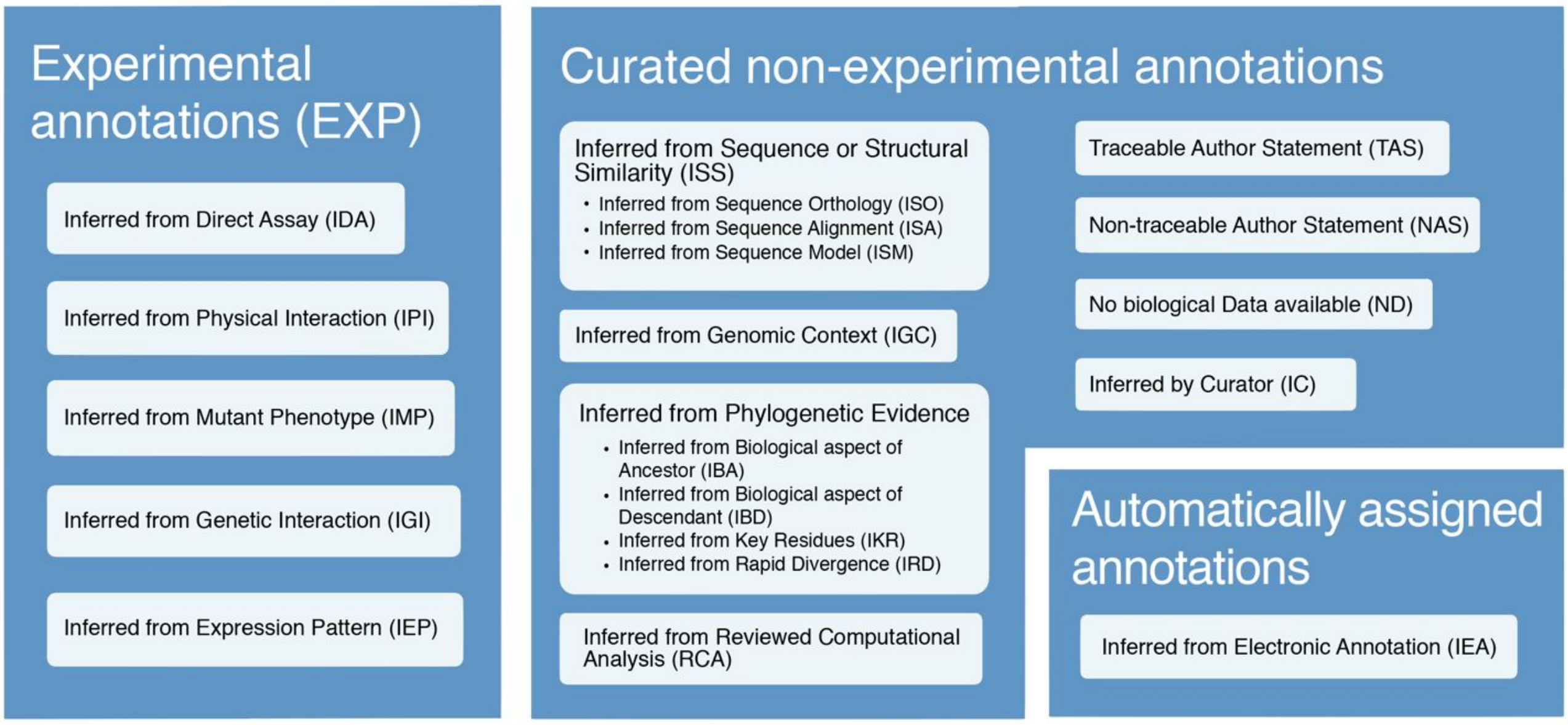

**Figure 3: GO Evidence Codes and their abbreviations.** The type of information supporting annotations is recorded with Evidence Codes, which can be grouped into three main categories: experimental evidence codes, curated non-experimental annotations, and automatically assigned annotations. The obsolete evidence code NR (Not Recorded) is not included in the figure. Documentation about the different types of automatically assigned annotations can be found at http://www.geneontology.org/doc/GO.references.

### 5.3.1 Experimentally supported annotations

Annotations based on direct experimental evidence found in the primary literature are denoted with the general evidence code EXP (Inferred from Experiment) or, when appropriate, the more specific evidence codes IDA (Inferred from Direct Assay), IPI (Inferred from Physical Interaction), IMP (Inferred from Mutant Phenotype), IGI (Inferred from Genetic Interaction), and IEP (Inferred from Expression Pattern) (Figure 3). These annotations are held in high regard by the community, e.g., ***(21)***, and are often used in applications such as checking the enrichment of a gene set in particular functions, finding genes that perform a specific function, or assessing involvement in specific specific pathways or processes.

Another important use of experimentally supported annotations is in providing trustworthy training sets for various computational methods that infer function ***(22)***. Used this way, the experimentally supported annotations can be amplified to understand more of the growing set of newly sequenced genes.

**5.3.2 Curated non-experimental annotations**

Fourteen of the twenty-one evidence codes are associated with manually curated non-experimental annotations. Annotations associated with these codes are curated in the sense that every annotation is reviewed by a curator, but they are non-experimental in the sense that there is no direct experimental evidence in the primary literature underpinning them; instead, they are inferred by curators based on different kinds of analyses.

ISS (Inferred from Sequence or Structural Similarity) is a superclass (i.e., a parent) of ISA (Inferred from Sequence Alignment), ISO (Inferred from Sequence Orthology), and ISM (Inferred from Sequence Model) evidence codes. Each of the he three sub-categories of ISS should be used when only one method was used to make the inference. For example, to improve the accuracy of function propagation by sequence similarity, many methods take into account the evolutionary relationships among genes. Most of these methods rely on orthology (ISO evidence code), because the function of orthologs tends to be more conserved across species than paralogs ***(23, 24)***. In a typical analysis, characterised and uncharacterised genes are clustered based on sequence similarity measures and phylogenetic relationships. The function of unknown genes is then inferred from the function of characterised genes within the same cluster ***(e.g. 25, 26)***.

Another approach to function prediction entails supervised machine learning based on features derived from protein sequence ***(27–30)*** (ISM evidence code). Such approach uses a training set of classified sequences to learn features that can be used to infer gene functions. Although few explicit assumptions about the complex relationship between protein sequence and function are required, the results are dependent on the accuracy and completeness of the training data.

IGC (Inferred from Genomic Context) includes, but is not limited to, such things as identity of the genes neighboring the gene product in question (i.e. synteny), operon structure, and phylogenetic or other whole genome analysis.

Relatively new are four evidence codes associated with phylogenetic analyses. IBA (Inferred from Biological aspect of Ancestor) and IBD (Inferred from Biological aspect of Descendant) indicate annotations that are propagated along a gene tree. Note that the latter is only applicable to ancestral genes. The loss of an active site, a binding site or a domain critical for a particular function can be annotated using the IKR (Inferred from Key Residues) evidence code. When this code is assigned by PAINT, GO's Phylogenetic Annotation and INference Tool ***(31)***, this means that it is a prediction based on evolutionary neighbors. Finally, negative annotations can be assigned to highly divergent sequences using the code IRD (Inferred from Rapid Divergence).

RCA (inferred from Reviewed Computational Analysis) captures annotations derived from predictions based on computational analyses of large-scale experimental data sets, or based on computational analyses that integrate datasets of several types, including experimental data (e.g. expression data, protein-protein interaction data, genetic interaction data, etc.), sequence data (e.g. promoter sequence, sequence-based structural predictions, etc.), or mathematical models.

Next, there are two types of annotations derived from author statements. Traceable Author Statement (TAS) refers to papers where the result is cited, but not the original evidence itself, such as review papers. On the other hand a NAS (Non-traceable Author Statement) refers to a statement in a database entry or statements in papers that cannot be traced to another paper.

The final two evidence codes for curated non-experimental annotations are IC (Inferred by Curator) and ND (No biological Data available). If an assignment of a GO term is made using the curator's expert knowledge, concluding from the context of the available data, but without any *direct* evidence available, the IC evidence code is used. For example, if a eukaryotic protein is annotated with the MF term "DNA ligase activity," the curator can assign the BP term "DNA ligation" and CC term "nucleus" with the evidence code IC.

The ND evidence code indicates that the function is currently unknown (i.e. that no characterization of the gene is currently available). Such an annotation is made to the root of the respective ontology to indicate which functional aspect is unknown. Hence, the ND evidence code allows users for a subtle difference between unannotated genes (for which the literature has not been completely reviewed and thus no GO annotation has been made) and uncharacterised genes (GO annotation with ND code). Note that the ND code is also different from an annotation with the "NOT" qualifier (which indicates the absence of a particular function).

### 5.3.3 Automatically assigned annotations

The evidence code IEA (Inferred from Electronic Annotation) is used for all inferences made without human supervision, regardless of the method used. IEA evidence code is by far the most abundantly used evidence code. The guiding idea behind computational function annotation is the notion that genes with similar sequences or structures are likely to be evolutionarily related, and thus, assuming they largely kept their ancestral function, they might still have similar functional roles today. For an in-depth discussion of computational methods for GO function annotations, refer to Chap. XX (chapter by Cozzetto and Jones) or see ***(32)***.

### 5.3.4 Additional considerations about evidence codes

Biases associated with the different evidence codes are discussed in the chapter by Gaudet and Dessimoz (x-ref). Note that there is a more extensive Evidence and Conclusion Ontology ***(ECO; 33)***, formerly known as the "Evidence Code Ontology", presented in Chap XXX. ECO is only partially implemented in the GO: ECOs are displayed in the AmiGO browser, but they are not in the GAF file. However, all Evidence Codes used by the GO are found also in ECO. There is a general assumption among the GO user community that annotations based on experiments are of higher quality compared to those generated electronically, but this has yet to be empirically demonstrated. Generally, annotations derived from automatic methods tend to be to high level terms, so they may have a lower information value, but they often withstand scrutiny. Conversely, experiments are sometimes overinterpreted (see Gaudet and Poux chapter) and can also contain inaccuracies.

## 5.4 Uniqueness of GO annotations (or lack thereof)

No two annotations can have the same combination of the following fields: gene/protein ID, GO term, evidence code, reference and isoform. Thus one gene can be annotated to the same term with more than one evidence code.

Most GO analyses are gene-based, and therefore it is important in such analyses to make sure the list of genes is non-redundant. However, annotations are often made to larger protein sets that include multiple proteins from the same gene. This is particularly evident in UniProt, which can contain distinct entries from the TrEMBL (unreviewed) portion of the database that do not necessarily represent biologically distinct proteins. The different entries for the same protein or gene are often annotated with identical GO terms, which can bias statistical analyses because some genes have many more entries than other genes. For instance, the set of human proteins in UniProt comprises over 70,000 entries, but there are only approximately 20,000 recognized human protein coding genes (20,187 reviewed human proteins in the UniProt release of 2015_12). The GO Consortium has worked with UniProt as well as the Quest for Orthologs Consortium to develop “gene-centric” reference proteome lists (http://www.uniprot.org/proteomes/) that provide a single “canonical” UniProt entry for each protein-coding gene. These lists are available for many species, and we encourage users performing gene-centric GO analyses to use only the annotations for UniProt entries in these lists.

## 6 How can I learn more about Gene Ontology resources?

Most of the topics introduced in this primer will be treated in more depth and nuance in later chapters. Part II focuses on the creation of GO function annotations—we cover in depth the two main strategies of creating GO function annotations: manual extraction/curation from the literature and computational prediction. Part III describes the main strategies used to evaluate their predictive performance. Part IV covers practical uses of the GO annotations: we discuss how GO terms and GO annotations can be summed and compared, how enrichment in specific GO terms can be analyzed, and how the GO annotations can be visualized. For the advanced GO user, part V discusses how the context of a GO annotation is recorded and goes beyond the Evidence Codes to describe how to capture more information on the source of an annotation. We end with part VI by going beyond GO: we present alternatives to GO for functional annotation; we show how a structured vocabulary is used in the context of controlled clinical terminologies; and we present how information from different structured vocabularies is integrated in one overarching resource.

## Acknowledgements

The authors gratefully acknowledge extensive feedback and ideas from Kimberly Van Auken, Marcus C. Chibucos, Prudence Mutowo, and Paul D. Thomas. PG acknowledges National

Institutes of Health/National Human Genome Research Institute grant HG002273. CD acknowledges Swiss National Science Foundation grant 150654 and UK BBSRC grant BB/M015009/1. JH acknowledges National Institutes of Health/National Institute for General Medical Sciences grant U24GM088849.